\documentclass[conference]{IEEEtran}
\IEEEoverridecommandlockouts
\usepackage{url}
\usepackage{graphicx}
\usepackage{amssymb}
\usepackage[english]{babel}
\usepackage{amsthm}
\theoremstyle{definition}
\usepackage{mathtools}
\usepackage{multirow}
\usepackage{multicol}
\usepackage{xcolor}
\usepackage{cite}
\usepackage{pifont}
\newcommand{\cmark}{\ding{51}}%
\newcommand{\xmark}{\ding{55}}%
\usepackage{multicol}
\usepackage[font=footnotesize]{subcaption}
\usepackage{longtable}
\usepackage{physics}
\usepackage{float}
\usepackage{cite}
\usepackage{float}
\usepackage{amsmath}
\usepackage{mathtools}
\usepackage{enumitem}
\usepackage{multirow}
\usepackage{balance}
\usepackage{multirow}
\newtheorem{theorem}{Property}

\usepackage[belowskip=-4pt,aboveskip=5pt]{caption}
\def\BibTeX{{\rm B\kern-.05em{\sc i\kern-.025em b}\kern-.08em
    T\kern-.1667em\lower.7ex\hbox{E}\kern-.125emX}}
\begin{document}

\title{Passivity of Electrical Transmission Networks modelled using  Rectangular and Polar D-Q variables}

\author{\IEEEauthorblockN{Kaustav Dey}
\IEEEauthorblockA{\textit{Department of Electrical Engineering} \\
\textit{Indian Institute of Technology Bombay}\\
Mumbai, India \\
email: kaustavd@iitb.ac.in}
\and
\IEEEauthorblockN{A. M. Kulkarni}
\IEEEauthorblockA{\textit{Department of Electrical Engineering} \\
\textit{Indian Institute of Technology Bombay}\\
Mumbai, India \\
email: anil@ee.iitb.ac.in} }

\maketitle

\begin{abstract}
The increasing penetration of converter-interfaced  distributed energy resources has brought out the need to develop decentralized criteria that would ensure the small-signal stability of the inter-connected system. Passivity of the D-Q admittance or impedance is a promising candidate for such an approach.  It is facilitated by the inherent passivity of the D-Q  impedance of an electrical network. However, the passivity conditions are generally restrictive and cannot be complied with in the low frequency range by the D-Q  admittance of devices that follow typical power control strategies. However, this does not imply that the system is unstable. Therefore, alternative formulations that use polar variables (magnitude/phase angle of voltages and real/reactive power injection instead of the D-Q components of voltages and currents) are  investigated. Passivity properties of the electrical network using these different formulations are brought out in this paper through analytical results and illustrative examples.

% passivity is a conservative criterion which is this is invariably violated in the low frequency range, when steady-state frequency and voltage regulation strategies are used. An alternative approach is to check if the passivity criterion can be satisfied in other input-output variables. To this end, this paper presents the passivity behaviour of the electrical transmission network, when different input-output variables as well as modelling simplifications are considered. It is observed tht although the dynamical model of the network is not passive in these alternative input-output variables, it can be passivated in some cases with voltage regulation capabilities of the shunt-connected devices. The analytical results presented in this paper are validated using numerical simulations on the transmission network model of the IEEE 9-bus system. 
\end{abstract}

\begin{IEEEkeywords}
Passive systems, small-signal stability, T\&D network passivity, Network Jacobian, Grid resonance.
\end{IEEEkeywords}

\section{Introduction}
A power system consists of a large number of devices like conventional and renewable energy generators, storage systems, FACTS and HVDC converters, which are connected to the Transmission \& Distribution (T\&D) network. Sometimes, adverse dynamic interactions between these devices and the network occur, leading to oscillatory instabilities~\cite{irwin2011ssci}. These instabilities can be analyzed with the help of stability assessment tools like time-domain simulation and eigenvalue analysis. These studies generally need to consider a large set of operating conditions of the network and the connected devices. Therefore, the possibility of specifying decentralized criteria which, if complied with by individual devices, would ensure the stability of the interconnected system, has great appeal. 
% Although local criteria are likely to be conservative (sufficient, but not necessary), this is acceptable as long as the criteria are not onerous or infeasible.
\par Passivity~\cite{khalil_non_linear} is a concept well-suited for this purpose because it is a sufficient stability criterion and is easy to evaluate in the frequency domain. Moreover, a T\&D network consisting of transmission lines, transformers, capacitors and inductors is inherently passive when it is formulated with currents and voltages as the interface (input or output)  variables. Hence the task is reduced to assessing passivity of the devices connected to the network.  The synchronously rotating (D-Q) coordinate system is convenient to do the passivity analysis as most devices are time-invariant in this frame of reference. Therefore, passivity of the D-Q based admittance of the shunt-connected devices has  been used in the past to prevent adverse device-grid  interactions~\cite{harnefors2015review,passivity_dc_microgrid,blaabjerg_active_damping_converters,harnefors2007input}. 
\par The passivity of the admittance of converters and synchronous machines with their controllers~(small-signal  models) have been analysed in~\cite{kaustav_jepes}. It has been found that the frequency domain passivity conditions are invariably violated in the low frequency range, when droop based frequency and voltage control strategies are used. However, this inherent non-passivity does not imply that the system will be unstable. Therefore, the passivity constraints on the D-Q based admittance are too restrictive for wide-band dynamic models of these devices. Therefore, it may be necessary to consider the high and low frequency models in a decoupled fashion (with the assumption that transients are time-scale separated), and formulate the low frequency device models with other input-output variables. These include the active and reactive power injections, and the  polar components of the bus voltage and their derivatives. However, whether the T\&D network retains its passivity with these formulations needs to be investigated.
\par With this motivation, this paper derives analytical results pertaining to the passivity behaviour of the T\&D network model when represented using the  alternative input-output variables.  It is found that the wide-band dynamical model of the T\&D network is not passive when represented using these alternative interface variables. 
% The restricted applicability of this scheme for slower transients is also investigated. 
Further, it is shown that the low frequency model of the T\&D network can be passivated if  shunt-connected devices to the network can provide voltage regulation capabilities. The results are verified with numerical simulations on the T\&D network of the IEEE 9-bus system~\cite{anderson2003power}.

\section{Passive Systems:Definition}
A dynamical system between the inputs $u(t)$ and an equal number of outputs $y(t)$ is called passive if it satisfies~\cite{khalil_non_linear}
\begin{align} \label{Eq:passive_storage_func}
u(t)^T y(t) \geq \dot{S}(x)
\end{align}
for all inputs and initial conditions. $S(x)$ is a continuously differentiable positive semi-definite function of the state variables $x(t)$, and is called the ``storage function'' of the system. The passivity definition for linear time-invariant~(LTI) systems is now presented.
\subsection{Passivity of LTI Systems (Time Domain Conditions)} \label{Sec:PR_ss}
A LTI system represented by the state space model ($A$, $B$, $C$, $D$), is passive if there exist matrices $P$, $Q$ and $W$ of appropriate dimensions such that~\cite{khalil_non_linear}
\begin{equation*}
    \begin{aligned} %\label{Eq:kyp_lemma}
P A + A^T P &= -Q^TQ, PB = C^T - Q^TW, D+D^T = W^TW
\end{aligned}
\end{equation*}
where the matrix $P$ is symmetric positive definite.
\subsection{Passivity of LTI Systems (Frequency Domain Conditions)} \label{Sec:PR_sys1}
\par A LTI system represented by a $n \times n$ rational, proper transfer function matrix $G(s)$ is passive if
\begin{enumerate}[leftmargin=*]
    \item there are no poles in the right half $s$-plane~(complex plane).
    \item The matrix $G^\mathcal{R}(j\Omega) = G(j\Omega) + G^T(-j\Omega)$ is positive semi-definite  for all $\Omega \in (-\infty, \infty)$ which is not a pole of $G(s)$. The superscript $T$ denotes the transpose operation.
    %\footnote{The superscripts $T$ and $H$ denote the transpose and conjugate-transpose operations respectively.}
    %$G^\mathcal{R}(j\Omega)$ is a Hermitian matrix, and therefore it will always have real eigenvalues. 
    %These should be non-negative for positive semi-definiteness~\cite{watkins2004fundamentals}.
    \item For all $j \Omega_p$ that are poles of $G(s)$, the poles must be simple and the residue of that pole $\lim_{s \to j\Omega_p} (s-j \Omega_p)\;G(s)$ should be positive semi-definite Hermitian.
\end{enumerate}
For LTI systems, the time domain and frequency domain conditions are equivalent. The usefulness of the passivity criterion for stability assessment stems from the following  properties.
\begin{enumerate}[label=\alph*),leftmargin=*]
    \item Passivity is a {\em sufficient} condition for stability.
     \item The inverse of a passive system is also passive, assuming that the state-space representation of the inverse system is well-defined.
    \item A system formed by a negative feedback connection of passive sub-systems is also passive.
\end{enumerate} 
The passivity of the small-signal model of the T\&D network, when represented using different input-output variables, is now presented.
\section{Model I variables: $(\Delta v_D, \Delta v_Q)$ and $(\Delta i_D, \Delta i_Q)$} \label{Sec:Model1}
In general, a T\&D network can be modelled as a multi-port admittance/impedance transfer function; the currents/voltages being the interface (input-output) variables.  The  storage  function for  this  system  can  be  chosen  to  be  the  electro-magnetic energy  stored  in  the  inductive  and  capacitive  components, which is dissipated in the resistive parts of these components. Topological  changes  in  the  network  (due  to  the  addition  or removal  of  lines),  or  changes  in  the  operating  conditions  do not affect the passivity of the network. 
\begin{figure}[h]
\centering
\includegraphics[width=0.47\textwidth]{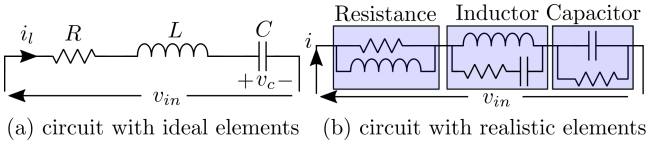}
\caption{Schematic of R-L-C series circuit}
\label{Fig:rlc_ckt}
\end{figure}
% 
%\begin{align*}
%    \frac{d}{dt} \begin{bmatrix}
%    i_l \\ v_c
%    \end{bmatrix} = \begin{bmatrix}
%    -\frac{R}{L} & -\frac{1}{L} \\ \frac{1}{C} & 0
%    \end{bmatrix} \begin{bmatrix}
%    i_l \\ v_c
%    \end{bmatrix} + \begin{bmatrix}
%    \frac{1}{L} \\ 0 
%    \end{bmatrix} v_{in}, \, i_l = \begin{bmatrix}
%    1 & 0
%    \end{bmatrix} \begin{bmatrix}
%    i_l \\ v_c
%    \end{bmatrix}
%\end{align*}
\par There are some subtleties though. While the admittance of the circuit in Fig.~\ref{Fig:rlc_ckt}(a) is passive, it is a strictly proper transfer function.
%\begin{align} \label{Eq:storage_func_rlc_example}
%S(x) = \frac{1}{2} \left( L i_l^2 + C v_c^2 \right)
%\end{align}
Therefore, the impedance transfer function $Z(s) = \frac{(s^2 LC + s RC + 1)}{sC}$ is not proper, and therefore, it is not passive. Non-properness  implies that $Z(s)\to \infty$ when $s\to \infty$, and that a connection to a current source  will lead to a singularity. However, in practice, the distributed nature of the circuit elements always alleviates such issues. For example, the consideration of the parasitic elements~(shown in Fig.~\ref{Fig:rlc_ckt}(b)) ensures that both the impedance and the admittance of the circuit are bi-proper. 
\par For a three-phase network, it is convenient to formulate the equations of the network in the D-Q frame. This is because most devices connected to the network are time-invariant in the D-Q domain, facilitating transfer function analyses. It is shown in~\cite{kaustav_jepes} that the passivity of the impedance/admittance of any component in three-phase  variables is retained under the synchronously rotating D-Q-o transformation. Therefore, the dynamical model of the T\&D network is also passive with $(\Delta v_D, \Delta v_Q)$ and $(\Delta i_D, \Delta i_Q)$ as the input-output variables\footnote{The zero sequence variables are generally stable, decoupled from the D-Q variables,  and localized to a small part of the network. Therefore, they are not considered in this analysis.}.
\begin{figure}[h]
    \centering
    \includegraphics[width=0.47\textwidth]{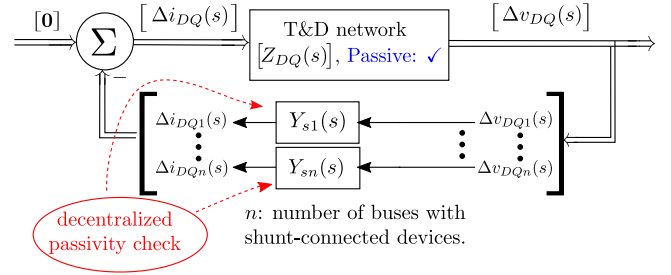}
    \caption{Scheme of passivity based stability criterion}
    \label{fig:decentralized_scheme}
\end{figure}
\par Since the admittance/impedance of a T\&D network is passive regardless of the  the topology or the operating condition, this can be exploited to develop a local small-signal stability assessment scheme, as shown in Figure~\ref{fig:decentralized_scheme}. In this scheme,  the D-Q domain admittance~(represented by $Y_{sj}(s)$ for device $j$ in Figure~\ref{fig:decentralized_scheme}) of the shunt-connected devices need to be individually passive in order to ensure the stability of the system. 
% \par \noindent \underline{Lossless Model:} It is a common approach to neglect the losses in the T\&D network model since they are very small. However, this may affect the properness of the impedance/admittance transfer functions. For example, the impedance transfer function  of the series R-L-C cicuit given in Figure~\ref{Fig:rlc_ckt}(b) is non-proper if the losses are neglected. However, in real-life, the parasitic elements will always take care of the properness issues. Therefore, the dynamical model of the D-Q domain impedance/impedance of lossless network~(along with the unmodelled parasitics) is also passive. 
 However, the violation of the frequency domain passivity conditions by  controlled power injection devices in the low frequency range restricts the  applicability of this scheme~\cite{kaustav_jepes}. To overcome this difficulty, the passivity  of these components, when modelled using  other input-output variables, is also examined. The choice of these alternative variables are motivated by the variables used in typical steady-state real and reactive power control strategies. 
% The interface variables in this case are the active and reactive power injections and the polar components of the bus voltage. 
\section{Model II variables: $(\Delta \phi, \Delta V_n)$ and $(\Delta P, \Delta Q)$}
The interface variables in this case are the active and reactive power injections and the polar components of the bus voltage. Let $P_j$ and $Q_j$ denote the active and reactive power injection at bus $j$ respectively, which are represented as follows.
\begin{align} \label{Eq:pq_idiq_vdvq}
P_j = v_{Dj} i_{Dj} + v_{Qj} i_{Qj}\, , \, Q_j = v_{Dj} i_{Qj} - v_{Qj} i_{Dj}
\end{align}
where $v_{Dj}, v_{Qj}, i_{Dj}, i_{Qj}$ represent the instantaneous D-Q components of the bus voltage and current injection at bus $j$. The bus voltage angle~$\phi_j$ and normalized magnitude~$V_{nj}$ of bus $j$ are related to $v_{Dj}$ and $v_{Qj}$ as follows.
$$
\phi_j = \tan^{-1} \left( \frac{v_{Dj}}{v_{Qj}} \right) ,\, V_{nj} = \sqrt{\frac{v_{Dj}^2 + v_{Qj}^2}{v_{Djo}^2 + v_{Qjo}^2}}
$$
where the subscript $o$ denotes the quiescent value of the corresponding variable. The subscript $j$ is dropped from the respective notations to denote the vector variables. For example, $P$ and $Q$ represent the vector of active and reactive power injections at all buses respectively. If there are $n$ buses, then 
\begin{align}
\begin{aligned} \label{Eq:vd_vq_id_iq_def}
v_{D} &= \text{diag}(v_{D1o}, .., v_{Dno}), v_{Q} = \text{diag}(v_{Q1o}, .., v_{Qno}) \\ i_{D} &= \text{diag}(i_{D1o}, .., i_{Dno}), i_{Q} = \text{diag}(i_{Q1o}, .., i_{Qno})
\end{aligned}
\end{align}
Note that diag($a,b$) represents a diagonal matrix with $a$ and $b$ as the diagonal entries. The equations of~\eqref{Eq:pq_idiq_vdvq} are linearized, and their small-signal variations are represented as follows.
\begin{align*}
\begin{bmatrix}
\Delta P(s) \\ \Delta Q(s)
\end{bmatrix} = \setlength\arraycolsep{1.6pt} \begin{bmatrix}
J_{11}(s) & J_{12}(s) \\ J_{21}(s) & J_{22}(s)
\end{bmatrix} \begin{bmatrix}
\Delta \phi(s) \\ \Delta V_n(s)
\end{bmatrix} = J(s) \begin{bmatrix}
\Delta \phi(s) \\ \Delta V_n(s)
\end{bmatrix}
\end{align*} 
$J(s)$ is related to the D-Q admittance~$Y_{DQ}(s)$ as follows.
\begin{align} 
J(s)  &= \left(  \mathcal{E}  Y_{DQ}(s)  +  \mathcal{C} \right)  \mathcal{F} \quad \text{where } \nonumber \\
 \mathcal{E}  = \setlength\arraycolsep{1.6pt} \begin{bmatrix}   v_{D}  &  v_{Q} \\ - v_{Q}  &  v_{D}  \end{bmatrix}, \, &\mathcal{C}  = \setlength\arraycolsep{1.6pt} \begin{bmatrix}  i_{D}  &  i_{Q} \\  i_{Q}  & -i_{D}  \end{bmatrix},  \mathcal{F}  = \setlength\arraycolsep{1.6pt} \begin{bmatrix}   v_{Q}  &  v_{D}  \\ -v_{D} &  v_{Q} \end{bmatrix} \label{Eq:vc_ic_vt}
\end{align}
%A state-space representation of $J(s)$ can also be represented in terms of the state-space matrices $(A_y,B_y,C_y,D_y)$, which represents the D-Q domain admittance $Y_{DQ}(s)$. The state-space representation $(A_2,B_2,C_2,D_2)$ of $J(s)$ satisfies the following relationship. 
%\begin{align} \label{Eq:ss_jacobian}
%A_2 = A_y,\, B_2 = B_y \mathcal{F}, C_2 = \mathcal{E} C_y, \, D_2 = (\mathcal{E} D_y + \mathcal{C}) \mathcal{F}
%\end{align} 
Note that $v_D, v_Q, i_D$ and $i_Q$ are defined in~\eqref{Eq:vd_vq_id_iq_def}. The following property describes the passivity behaviour of the dynamical model of the T\&D network when $(\Delta \phi, \Delta V_n)$ and $(\Delta P, \Delta Q)$ are used as the interface variables. 
\subsection{Wide-band dynamical model}

\begin{theorem} \label{Th:rlc_jacobian_non_passive}
The dynamical model of a R-L-C network with $(\Delta P,  \Delta Q)-(\Delta \phi,  \Delta V_n)$ as interface variables is not passive. 
\end{theorem}
\noindent The proof is given in Appendix~\ref{AppSec:passivity_network_dynamic_model}-1. Although this restricts the applicability of passivity based scheme using wide-band models, a limited scheme of applicability for the slower transients using low frequency models is considered. The passivity of the low frequency model is now presented.
% Note that the proof does not make any assumptions about the losses in the network. Therefore, Property~\ref{Th:rlc_jacobian_non_passive} holds for the dynamical models of both lossy and lossless networks.
\subsection{Low frequency model}
For typical parameters, the natural modes of the T\&D network model~(in D-Q variables) usually lie in the high frequency range. Therefore, it is reasonable to represent the network by a static low-frequency model for slower transient studies. At low frequencies, the network transfer function can be approximated by $J_{LF}$, which is the unreduced form of the load-flow Jacobian matrix. 
\begin{align}
J_{LF} = J(0) = \begin{bmatrix}
J_{LF11} & J_{LF12} \\ J_{LF21} & J_{LF22}
\end{bmatrix} = \begin{bmatrix}
J_{11}(0) & J_{12}(0) \\ J_{21}(0) & J_{22}(0)
\end{bmatrix}
\end{align}
Note that $J_{LF}$ represents a static relation. Normally $J_{LF}$ is singular~(with one zero eigenvalue), because there cannot be a unique solution for the phase angles for specified set of active and reactive power injections. $J_{LF}^{\mathcal{R}} = J_{LF} + J_{LF}^T$ may not be  positive semi-definite because the shunt capacitances in the network often result in a negative eigenvalue. This implies that $J_{LF}$ may not be passive. 
\par
\begin{figure}[h]
    \centering
    \includegraphics[width=0.44\textwidth]{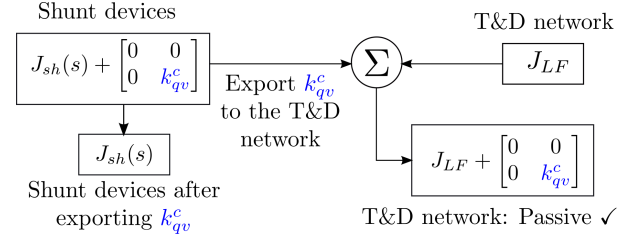}
    \caption{Contribution of devices to network passivity}
    \label{fig:kqv_export_scheme}
\end{figure}

 The non-passivity of $J_{LF}$ can be alleviated by {\em requiring} that some devices connected to the network ``contribute'' to the diagonal terms of $J_{LF22}$. This is in order to compensate the effect of the shunt capacitances. In practical terms, this means that at least some  devices should contribute to voltage regulation through $\Delta Q- \Delta V_n$ droop control~(``grid-forming'' devices), as shown in Figure~\ref{fig:kqv_export_scheme}. The contribution, denoted by $k_{qv}^{c}$, has to be specified for each device by the T\&D operator based on an evaluation of the eigenvalues of $J_{LF}^R$. The aim is to ensure that no eigenvalue of $J_{LF}^R$ is negative. 

\par
\begin{figure}[h]
    \centering
    \includegraphics[width=0.45\textwidth]{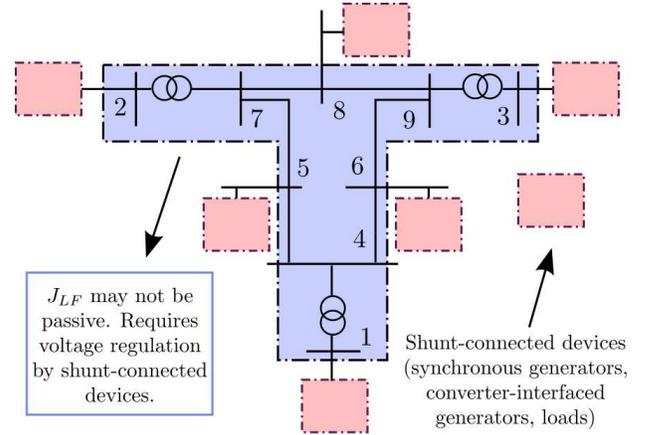}
    \caption{A network with controllable devices}
    \label{fig:three_mac_system_schematic}
\end{figure}

 \noindent \underline{Illustrative Example:} The schematic of a network with controllable generators and loads is shown in Figure~\ref{fig:three_mac_system_schematic}. The transmission line parameters and the equilibrium power flows are 
 taken from the three-machine system  of~\cite{anderson2003power}. 
\par The eigenvalues of $J_{LF}^{\mathcal{R}}$ are given in Table~\ref{Tab:three_mac_sys_network_jacobian_eig}. $J_{LF}$ is not passive as $J_{LF}^{\mathcal{R}}$ has a negative eigenvalue. If all the controllable generators and loads are able to contribute $\Delta Q = 0.65 \Delta V_n$~(in pu), then $J_{LF}$ becomes passive, as shown in the table. 
% Although the passivation of $J_{LF}$ is demonstrated here with equal contribution of voltage regulation by all devices, it need not always be so in practice.
\begin{table}[h]
\caption{Eigenvalues of $J_{LF}^{\mathcal{R}}$ of the three machine system}
\label{Tab:three_mac_sys_network_jacobian_eig}
\begin{tabular}{|c|c|}
\hline
Base Case &
  Modified Case \\ \hline
\begin{tabular}[c]{@{}c@{}}$\mathbf{-0.84}, 0, 7.52, 8.50, 10.15,$ \\ $ 12.93, 31.71, 34.74, 34.95,$ \\ $36.29, 41.37, 42.8, 93.51, 94.52,$\\ $107.94, 108.29, 115.46, 115.76$\end{tabular} &
  \begin{tabular}[c]{@{}c@{}}$0, \mathbf{0.025}, 7.87, 8.82, 10.42,$\\ $13.13, 32.43, 35.06, 35.44, 36.53,$\\ $42.1, 42.88, 93.92, 95.08,$\\ $ 108.29, 109.06, 115.72, 116.61$\end{tabular} \\ \hline
\multicolumn{2}{|c|}{Reactive power-voltage regulation: 0.65 pu at buses 1, 2, 3, 5, 6, 8 each.} \\ \hline
\end{tabular}
\end{table}
\par \noindent \underline{Lossless Model:} The lossless approximation does not alleviate the non-passivity of $J_{LF}$. However, similar to the previous case, the non-passivity can be alleviated by contribution of voltage regulation by the shunt-connected devices.
\par This is also numerically verified using the three-machine system shown in Figure~\ref{fig:three_mac_system_schematic}. The line resistances are neglected, and then $J_{LF}$ is calculated. The eigenvalues of $J_{LF}^{\mathcal{R}}$, with and without shunt voltage regulation contributions, are presented in Table~\ref{Tab:three_mac_sys_network_jacobian_eig_lossless}. Therefore, the low-frequency lossless T\&D network model can also be passivated by borrowing voltage regulation capabilities from the shunt-connected devices.
\begin{table}[h]
\caption{Eigenvalues of $J_{LF}^{\mathcal{R}}$ of the three machine system}
\label{Tab:three_mac_sys_network_jacobian_eig_lossless}
\begin{tabular}{|c|c|}
\hline
Base Case &
  Modified Case \\ \hline
\begin{tabular}[c]{@{}c@{}}$\mathbf{-0.84}, 0, 7.8, 8.83, 10.32,$ \\ $ 13.11, 32.72, 35.46, 35.74,$ \\ $36.79, 41.72, 43.16, 94.37, 95.34,$\\ $109, 109.33, 116.92, 117.26$\end{tabular} &
  \begin{tabular}[c]{@{}c@{}}$0, \mathbf{0.027}, 8.15, 9.15, 10.59,$\\ $13.31, 33.44, 35.86, 36.11, 37.07,$\\ $42.45, 43.25, 94.77, 95.91,$\\ $ 109.33, 110.1, 117.2, 118.09$\end{tabular} \\ \hline
\multicolumn{2}{|c|}{Reactive power-voltage regulation: 0.65 pu at buses 1, 2, 3, 5, 6, 8 each.} \\ \hline
\end{tabular}
\end{table}
\par \noindent \underline{Decoupled Model $(J_{LF12} = J_{LF21} = 0)$:} For transmission network parameters, the off-diagonal blocks of $J_{LF}$ are usually much smaller than the diagonal blocks. Therefore, a further simplified model is also considered where  $J_{LF12} = J_{LF21} = 0$. This will be referred to as the ``decoupled'' model in this paper. The passivity behaviour is as follows.\\
(a) Decoupled lossy network: $J_{LF}$ is not passive.\\
(b) Decoupled lossless network: $J_{LF}$ is not passive if shunt capacitances are considered. \\
However, for both (a) and (b), voltage regulation by the shunt-connected devices can alleviate the non-passivity.\\
(c) $J_{LF}$ is passive in the case of decoupled lossless model, with shunt capacitances also neglected.
\section{Model III variables: $(\Delta P, \Delta Q)$ -- $(\Delta \tilde{\omega}, \Delta V_n)$}
In contrast to the input-output variables used in the previous case, the bus frequency deviation~($\tilde{\omega}$) is considered here instead of the bus phase angle. The derivative is approximated  by using a small time-constant~($\tau$) as given in~\eqref{Eq:omega_derv_approx}.
\begin{align} \label{Eq:omega_derv_approx}
\Delta \tilde{\omega}(s) = \frac{s}{(1+s \tau)} \Delta \phi(s)
\end{align}
Let the transfer function of the T\&D network in these variables be denoted by $J_{dp}(s)$. This is related to $J(s)$ as follows.
\begin{align}
J_{dp}(s) = \begin{bmatrix}
J_{11}(s) & J_{12}(s) \\ J_{21}(s) & J_{22}(s)
\end{bmatrix} \begin{bmatrix}
\frac{(1+s \tau)}{s} I & 0 \\ 0 & I
\end{bmatrix}
\end{align}
where $I$ is the identity matrix. The following property reflects the passivity behaviour of the dynamical model of the T\&D network, when these input-output variables are considered.
\subsection{Wide-band dynamical model}
\begin{theorem} \label{Th:rlc_jacobian_f_V}
The dynamical model of a R-L-C network with $(\Delta P, \Delta Q)-(\Delta \tilde{\omega},  \Delta V_n)$ as interface variables is not passive. 
\end{theorem}
The proof is given in Appendix~\ref{AppSec:passivity_network_dynamic_model}-2. The passivity behaviour of the low frequency model is now presented. 
%The proof does not make any assumptions about the losses of the network. Therefore, Property~\ref{Th:rlc_jacobian_f_V} is valid for the dynamical models of both lossless and lossy networks.
\subsection{Low frequency model}
The low frequency transfer function $N_{p}(s)$ is given in~\eqref{Eq:R_partial_der}.
\begin{align} \label{Eq:R_partial_der}
N_{p}(s) =  J(0) \begin{bmatrix}
\frac{(1+s \tau)}{s} I & 0 \\ 0 & I
\end{bmatrix} = \begin{bmatrix}
\frac{J_{LF11}(1+s \tau)}{s} & J_{LF12} \\ \frac{J_{LF21}(1+s \tau)}{s} & J_{LF22}
\end{bmatrix}
\end{align}
Note that $N_{p}(s)$ has a simple pole at $s = 0$. For the system to be passive, the residue at $s =0$ must be positive semi-definite Hermitian. The expression of the residue of $N_{p}(s)$ evaluated at $s = 0$, denoted by $S_{dp}\,$, is as follows.
\begin{align} \label{Eq:residue_model_3}
S_{dp} = \lim_{s \to 0} s N_{p}(s) = \begin{bmatrix}
J_{LF11} & 0 \\ J_{LF21} & 0
\end{bmatrix}
\end{align} 
Note that $S_{dp}$ cannot be Hermitian if $J_{LF12} \neq 0$. Therefore, the low frequency model of the network in these variables cannot be passive, if coupled power flow models are considered.
\par \noindent \underline{Decoupled model:} $S_{dp}$ in~\eqref{Eq:residue_model_3} is positive semi-definite  if $J_{LF11}$ is Hermitian positive semi-definite. Note that  $J_{LF11}$ is Hermitian only for lossless networks. Therefore, $N_{dp}(s)$ of decoupled lossy T\&D network models is also not passive.
\par \noindent \underline{Lossless network:} For decoupled lossless networks at $\Omega \neq 0$,
\begin{align*}
N^{\mathcal{R}}_{dp}(j\Omega) = N_{p}(j\Omega) + N_{p}^T(-j\Omega) = \begin{bmatrix}
0 & 0 \\ 0 & J_{LF22} + J_{LF22}^T
\end{bmatrix}
\end{align*}
This may not be passive due to the presence of shunt capacitors. However, the non-passivity can be alleviated by contribution of voltage regulation by the shunt-connected devices. If the shunt capacitors are also neglected, then the network model is passive in these variables. This network model is used for passivity based stability analysis in~\cite{watson_ilc_passivity}.

\begin{table*}[h]
\caption{Passivity of T\&D network transfer function with different input-output variables}
\label{Tab:passivity_network_model_summary}
\begin{center}
\begin{tabular}{|c|c|c|c|c|cccc|}
\hline
\multirow{2}{*}{\begin{tabular}[c]{@{}c@{}}Model \end{tabular}} & 
\multirow{2}{*}{\begin{tabular}[c]{@{}c@{}}Inputs \end{tabular}} & \multirow{2}{*}{\begin{tabular}[c]{@{}c@{}}Outputs \end{tabular}} & \multirow{2}{*}{\begin{tabular}[c]{@{}c@{}}Transfer\\ function \end{tabular}} & \multirow{2}{*}{\begin{tabular}[c]{@{}c@{}}Wide-band \\ dynamic model \end{tabular}} & \multicolumn{4}{c|}{Low-frequency model} \\ \cline{6-9} 
& & & & \multicolumn{1}{c|}{\begin{tabular}[c]{@{}c@{}}  \end{tabular}} & \multicolumn{1}{c|}{\begin{tabular}[c]{@{}c@{}}Lossy\\  (with $B$)\end{tabular}} & \multicolumn{1}{c|}{\begin{tabular}[c]{@{}c@{}}Lossless\\ (with $B$)\end{tabular}} & \multicolumn{1}{c|}{\begin{tabular}[c]{@{}c@{}}Lossy\\ (without $B$)\end{tabular}} & \begin{tabular}[c]{@{}c@{}}Lossless\\ (without $B$)\end{tabular} \\ \hline
Model I & $(\Delta i_D, \Delta i_Q)$ & $(\Delta v_D, \Delta v_Q)$ & $Z_{DQ}(s)$ & \cmark & \multicolumn{1}{c|}{\cmark} & \multicolumn{1}{c|}{\cmark} & \multicolumn{1}{c|}{\cmark} & \cmark \\ \hline
Model II & $(\Delta \phi, \Delta V_n)$ & $(\Delta P, \Delta Q)$ & $J(s)$ &   \xmark & \multicolumn{1}{c|}{\begin{tabular}[c]{@{}c@{}}Coupled: \xmark$^*$\\ Decoupled: \xmark$^*$\end{tabular}} & \multicolumn{1}{c|}{\begin{tabular}[c]{@{}c@{}}Coupled: \xmark$^*$\\ Decoupled: \xmark$^*$\end{tabular}} & \multicolumn{1}{c|}{\begin{tabular}[c]{@{}c@{}}Coupled: \xmark$^*$\\ Decoupled: \xmark$^*$\end{tabular}} & \begin{tabular}[c]{@{}c@{}}Coupled: \xmark$^*$\\ Decoupled: \cmark\end{tabular} \\ \hline
Model III & $(\Delta \tilde{\omega}, \Delta V_n)$ & $(\Delta P, \Delta Q)$ & $J_{dp}(s)$ & \xmark & \multicolumn{1}{c|}{\begin{tabular}[c]{@{}c@{}}Coupled: \xmark \\ Decoupled: \xmark\end{tabular}} & \multicolumn{1}{c|}{\begin{tabular}[c]{@{}c@{}}Coupled: \xmark\\ Decoupled: \xmark$^*$\end{tabular}} & \multicolumn{1}{c|}{\begin{tabular}[c]{@{}c@{}}Coupled: \xmark\\ Decoupled: \xmark\end{tabular}} & \begin{tabular}[c]{@{}c@{}}Coupled: \xmark\\ Decoupled: \cmark\end{tabular} \\ \hline
Model IV & $(\Delta \tilde{\omega}, \Delta \tilde{V}_n^d)$ & $(\Delta P, \Delta Q)$  & $J_{df}(s)$ & \xmark & \multicolumn{1}{c|}{\begin{tabular}[c]{@{}c@{}}Coupled: \xmark \\ Decoupled: \xmark\end{tabular}} & \multicolumn{1}{c|}{\begin{tabular}[c]{@{}c@{}}Coupled: \xmark$^*$\\ Decoupled: \xmark$^*$ \end{tabular}} & \multicolumn{1}{c|}{\begin{tabular}[c]{@{}c@{}}Coupled: \xmark \\ Decoupled: \xmark \end{tabular}} & \begin{tabular}[c]{@{}c@{}}Coupled: \xmark$^*$ \\ Decoupled: \cmark\end{tabular} \\ \hline
\multicolumn{9}{|c|}{*: Non-passivity can be alleviated by borrowing voltage regulation capability of shunt-connected devices. $B$ denotes shunt capacitance.} \\ \hline
\end{tabular}
\end{center}
\end{table*}

\section{Model IV variables: $(\Delta P, \Delta Q)$ -- $(\Delta \tilde{\omega}, \Delta \tilde{V}_n^d)$} \label{Sec:Model4}
The input variables that are considered here are $\Delta \tilde{\omega}$ and the derivative of $\Delta V_n$, denoted by $\Delta \tilde{V}_n^d\,$. Similar to the evaluation of $\tilde{\omega}$ as given in~\eqref{Eq:omega_derv_approx}, the derivative here is also approximated by the same time-constant $\tau$, as given in~\eqref{Eq:vn_dot}.
\begin{align} \label{Eq:vn_dot}
\Delta \tilde{V}_n^d(s) = \frac{s}{(1+s \tau)} \Delta V_n(s)
\end{align}
The transfer function in these variables $J_{df}(s)$ is related to $J(s)$ as follows.
\begin{align}
J_{df}(s) =  J(s) \times  \frac{(1+s \tau)}{s}
\end{align}
The following property reflects the passivity behaviour of the dynamical model of the T\&D network, when these input-output variables are considered.
%The state-space representation $(A_4, B_4, C_4, D_4)$ of $J_{df}(s)$ can be representated in terms of the state-space representation of $Y_{DQ}(s)$ as follows.
%\begin{align}
%A_4 = \frac{1}{\tau} \, \begin{bmatrix}
%A_2 & 0 \\ C_2 & -I
%\end{bmatrix}, \, B_4 = \begin{bmatrix}
%B_2 \\ \frac{D_2}{\tau}
%\end{bmatrix}, \, C_4^T =\frac{1}{\tau} \, \begin{bmatrix}
%C_2^T \\ -I
%\end{bmatrix}, \, D_4 = \frac{D_2}{\tau}
%\end{align}
%where $I$ denotes the identity matrix of appropriate dimensions and $(A_2, B_2, C_2, D_2)$ represents the state-space representation of $J(s)$, as given in~\eqref{Eq:ss_jacobian}.
\subsection{Wide-band dynamical model}
\begin{theorem} \label{Th:rlc_jacobian_der_non_passive}
The dynamical model of a R-L-C network with $(\Delta P$, $ \Delta Q)-(\Delta \tilde{\omega}$, $ \Delta \tilde{V}_n^d)$ as interface variables is not passive.
\end{theorem}
The proof is given in Appendix~\ref{AppSec:passivity_network_dynamic_model}-3. The passivity behaviour of the low frequency model is now presented. 
%Since the proof does not make any assumptions about the losses in the network, Property~\ref{Th:rlc_jacobian_der_non_passive} therefore holds true for the dynamical models of both lossy and lossless networks. 
\subsection{Low frequency model} The low frequency model of the T\&D network in these variables is represented by $N_{df}(s)$, which is given as follows.
\begin{align}
N_{df}(s) = J_{LF} \times \frac{(1+s\tau)}{s}
\end{align}
$N_{df}(s)$ has a simple pole at $s=0$. For $N_{df}(s)$ to be passive, the residue, evaluated at $s = 0$ has to be positive semi-definite Hermitian. The residue is $J_{LF}$, which  is not Hermitian for lossy networks. Therefore, the low frequency model of lossy networks will not be passive in these input-output variables.
% \begin{align}
% S_{df} = \lim_{s \to 0} s R_{df}(s) = R
% \end{align}

\par \noindent \underline{Lossless network:} For lossless networks, $N_{df}$ is passive if and only if $J_{LF}$ is positive semi-definite. Although $J_{LF}$ may not be positive semi-definite with shunt capacitances considered, it can be achieved with voltage regulation contribution by the shunt-connected devices. If the shunt capacitances are neglected and decoupling is considered, then $J_{LF}$ is positive semi-definite Hermitian.

%\newpage
%\vspace*{5mm}
%\newpage

%\section{Summary and Remarks}
% The following points are worth mentioning.\\
%(i) In the preceding sections, the passivity behaviour of the test system is numerically verified only for the Model II variables~(low frequency models). This is not explicitly shown for the other variable sets since the conclusions in these cases are derived from the conclusions of Model II variables. \\
%(ii)   

\section{Conclusions and Future Work}
The passivity behaviour of the T\&D network transfer function for D-Q based rectangular and polar~(and their derivatives) interface variables are summarized in Table~\ref{Tab:passivity_network_model_summary}.  The wide-band dynamic model of the impedance/admittance of the T\&D network is passive. Although the wide-band models of the T\&D network is not passive for the alternative interface variables considered here, the low-frequency models can be passivated by borrowing voltage regulation contributions from the shunt-connected devices. It is therefore necessary to ensure time-scale separation for decoupling the high and low frequency analysis in order to make the scheme viable.  The decoupled analysis can use the rectangular variables for high frequency studies, and the polar variables for low frequency studies. 
%It is expected that a passivity based local stability criterion can be developed encompassing both faster and slower transients, which is not too onerous to be met. 
% The passivity behaviour of the low frequency models can be used to locally assess the stability of the slower transients.
\bibliographystyle{IEEEtran}
\bibliography{reference}

\appendices

\renewcommand{\thesectiondis}[2]{\Alph{section}:}

\renewcommand{\thesubsectiondis}{\arabic{subsection}.}

\section{Passivity of T\&D network model} \label{AppSec:passivity_network_dynamic_model}
Let the state space matrices of the D-Q domain admittance $Y_{DQ}(s)$ be $(A_y, B_y, C_y, D_y)$. Then
\begin{align}
 D_y = \lim_{s \to \infty}  Y_{DQ}(s) = \begin{bmatrix}
 D_1  & \mathbf{0}  \\ \mathbf{0} & D_1
 \end{bmatrix}
%  \left[ \begin{array}{c|c} D_1  & \mathbf{0}  \\ \hline \mathbf{0} & D_1 \end{array} \right]
\end{align}
Note that $D_1$ is symmetric. 
\subsection{Model II variables: $(\Delta P, \Delta Q)$ -- $(\Delta \phi, \Delta V_n)$} \label{AppSec:subsec_rlc_pass_jacobian}
If the state-space matrices of the transfer function $J(s)$ are $(A_2, B_2,C_2,D_2)$, then $D_2 =  \left( \mathcal{E} D_y + \mathcal{C} \right) \mathcal{F}$,
%\begin{equation} \label{Eq:dj_jacobian_rlc}
%D_2 =  \left( \mathcal{E} D_n + \mathcal{C} \right) \mathcal{F} 
%\end{equation}
where $\mathcal{E}, \mathcal{C}$ and $\mathcal{F}$ are defined in~\eqref{Eq:vc_ic_vt}. 
%$$D_j + D_j^T = 2\begin{bmatrix}
% (i_{Do}v_{Qo}-i_{Qo} v_{Do}) &  (i_{Do}v_{Do}+i_{Qo} v_{Qo}) \\ (i_{Do}v_{Do}+i_{Qo} v_{Qo}) & - (i_{Do}v_{Qo}-i_{Qo} v_{Do}) 
%\end{bmatrix}$$ 
It can be shown that the trace of $D_2 + D_2^T$ is zero, indicating that it cannot be positive semi-definite. This violates the time domain conditions~(see Section~\ref{Sec:PR_ss}). Therefore the system is not passive.
\subsection{Model III variables: $(\Delta P, \Delta Q)$ -- $(\Delta \tilde{\omega}, \Delta V_n)$}
\label{AppSec:subsec_rlc_pass_jacobian_partial_der}
If the state-space matrices of the transfer function $J_{dp}(s)$ are $(A_{3}, B_{3},C_{3}, D_{3})$, then 
$$ D_3 = \left( \mathcal{E} D_y + \mathcal{C} \right) \mathcal{F} \begin{bmatrix}
\tau I & 0 \\ 0 & I
\end{bmatrix}  $$
where $I$ is the identity matrix. The diagonal entries of $D_3+D_3^T$ cannot be non-negative except when $i_{Do} v_{Qo} - i_{Qo} v_{Do} = 0$, which indicates that the quiescent reactive power injection at each node is zero, which is quite unusual. If  $i_{Do} v_{Qo} - i_{Qo} v_{Do} \neq 0$, then $D_3 + D_3^T$ cannot be positive semi-definite. This  will violate the time-domain passivity conditions~(see Section~\ref{Sec:PR_ss}). Therefore, the system is not passive if $Q_o \neq 0$.

\subsection{Model IV variables: $(\Delta P, \Delta Q)$ -- $(\Delta \tilde{\omega}, \Delta \tilde{V}_n^d)$}
\label{AppSec:subsec_rlc_pass_jacobian_der}
If the state-space matrices of the transfer function matrix $J_{df}(s)$ are $(A_{4}, B_{4},C_{4}, D_{4})$, then $ D_{4}  =   \tau \left( \mathcal{E} D_y + \mathcal{C} \right) \mathcal{F} $, where $\mathcal{E}, \mathcal{C}$ and $\mathcal{F}$ are as given in~\eqref{Eq:vc_ic_vt}. Note that the trace of $D_{4}+D_{4}^T$ is zero, indicating that it cannot be positive semi-definite. This violates the time domain passivity conditions~(see Section~\ref{Sec:PR_ss}). Therefore, it is not passive.

\end{document}